\documentclass[reqno,12pt]{article}
\usepackage{palatino}
\usepackage{bm}
\usepackage[utf8]{inputenc}
\usepackage[T1]{fontenc}
\usepackage{amsfonts}
\usepackage{latexsym}
\usepackage{dsfont}
\usepackage{amssymb,latexsym}
\usepackage{amsthm}
\usepackage[centertags]{amsmath}
\usepackage[french,english]{babel}
\usepackage{indentfirst}
\usepackage{graphicx}
\usepackage{hyperref}
\usepackage{geometry}
\theoremstyle{plain}

\theoremstyle{definition}

\theoremstyle{remark}

\begin{document}


\title{Applying k-nearest neighbors to time series forecasting: two new approaches}

\author{\textbf{S.Tajmouati $^{1,a}$; B.Wahbi$^1$; A.Bedoui $^2$;A.Abarda$^3$  and M. Dakkon $^4$}}
\date{$^1$Department of Mathematics, Ibn Tofail
University,Faculty of Sciences,
Kenitra, Morocco,\\ samya.tajmouati@gmail.com \\[0.1cm]
$^2$ Department of Statistics, University of Georgia, Athens GA, USA,\\
bedoui.adel1@gmail.com \\[0.1cm]
$^3$L.M.M.C.E, Universite Hassan 1er,FSJES,
Settat, Morocco \\
abdallah.abarda@uit.ac.ma \\[0.1cm]
$^4$The research team of Modeling and Information Theory, Université Abdelmalek Essaadi,FSJES \\ Tetouan, Morocco.\\
m.dakkoun@gmail.com\\[0.3cm]
 \small $^a$Corresponding author: S.Tajmouati
   \small (Email: samya.tajmouati@gmail.com)}

\maketitle

\begin{abstract}
\noindent K-nearest neighbors algorithm is one of the prominent techniques used in classification and regression. Despite its simplicity, the k-nearest neighbors has been successfully applied in time series forecasting. However, the selection of the number of neighbors and feature selection is a daunting task. In this paper, we introduce two methodologies to forecasting time series that we refer to as Classical Parameters Tuning in Weighted Nearest Neighbors and Fast Parameters Tuning in Weighted Nearest Neighbors. The first approach uses classical parameters tuning that compares the most recent subsequence with every possible subsequence from the past of the same length. The second approach reduces the neighbors' search set, which leads to significantly reduced grid size and hence a lower computational time. To tune the models' parameters, both methods implement an approach inspired by cross-validation for weighted nearest neighbors. We evaluate the forecasting performance and accuracy of our models. Then, we compare them to some classical approaches, especially, Seasonal Autoregressive Integrated Moving Average, Holt-Winters and Exponential Smoothing State Space Model. Real data examples on retail and food services sales in the USA and milk production in the UK are analyzed to demonstrate the application and the efficiency of the proposed approaches.

\end{abstract}


\section{Introduction}

 k-nearest neighbors (k-NN) algorithm is one of the most popular non-parametric approaches used for classification, and it has been extended to regression (Martínez et al., 2018). k-NN is a simple algorithm that has been effectively used in various research areas such as financial modeling (Zhang et al., 2017), image interpolation (Ni and Nguyen, 2009)  , and visual recognition (Liu, 2016). The k-NN algorithm has been successfully applied in time series forecasting despite its simplicity. For example, Tang, et al. (2018) propose an integrated k-NN regression model with principal component analysis (PCA) for financial time series prediction. That is, the historical data  are transformed by implementing the PCA method such that the resulting components that capture the largest amount of information are used as inputs in the k-NN model.  This approach speeds up the calculation process in the k-NN method and eliminates the non-relevant information.\\
Zhang, et al. (2017) use a multidimensional k-NN model combined with an ensemble empirical mode decomposition to predict the stock price. Martínez et al. (2019) propose a general methodology based on the k-NN approach to accurately forecast the N3 competition data set. Their objective is  to devise an automatic tool that selects the modeling strategies and preprocessing approaches for the N3 competition data set. Then, they propose a final method to forecast the time series, which is based on the results obtained from the Wilcoxon signed-rank test.  However, one should note that this approach might lead to different results pending on the order of the selected strategies.  In the field of electricity, some researchers implement the k-NN regression approach to estimate electricity demand and price.  Alvarez, et al. (2010) introduce a pattern sequence-based forecasting (PSF) algorithm based on the k-NN method to forecast electricity prices and demand.  First, a clustering technique is implemented to classify and label the days that constitute the series. Then, the labeled sequence that precedes the next day is used to predict both the price and demand by implementing the k-NN algorithm. Lora, et al. (2007) implement a weighted k-NN regression to forecast the electricity market price for the next day. Ahmed, et al. (2010) provide a comparative study of performance estimation methods based on machine learning algorithms for time series forecasting, including the k-NN regression. However, only one step ahead predictions are performed, and there is no study on the parameter selection like input variables, which is the objective of this paper.\\
The selection of the number of neighbors and performing feature selection in the k-NN approach for time series data are a daunting task.  A small alteration in values of the number of neighbors or feature selection reduces the model performance.  In this context, we introduce two automatic methods that we refer to as Classical Parameters Tuning in Weighted Nearest Neighbors (CPTO-WNN) and Fast Parameters Tuning in Weighted Nearest Neighbors (FPTO-WNN). The CPTO-WNN method compares the most recent subsequence with every possible subsequence from the past of the same length. After that, the algorithm selects the closest subsequences to the most recent subsequence. Then, we can forecast the future's subsequence  based on their next values. Although, searching every possible subsequence and testing its closeness to the most recent subsequence may lead to efficient results, it will certainly raise concerns about the computational time. To overcome this issue, we introduce the FPTO-WNN method that lowers the computational cost in the CPTO-WNN method by reducing the neighbors' search set. Both methods implement an approach inspired by the cross-validation method that selects an optimal number of neighbors and feature selection. That is, the data are divided into  training sets  and testing sets. The number of neighbors and features are selected by minimizing the average accuracy measure, which is calculated over all the test sets.\\
The outline of the remaining sections is as follows.  In Section 2, we demonstrate the application of k-NN regression to time series forecasting. Section 3 describes different preprocessing techniques based on the k-NN regression setting. In Section 4, we present our proposed methods. Two real data analyses and a comparison study are presented in Sections 5 and 6, respectively. Section 7 contains conclusions.


\section{Time series forecasting using k-NN approach}

 k-NN is one of the simplest classical algorithms used in machine learning. It was first used in classification. For unlabeled example, the k-NN approach searches among all the labeled cases the k closest examples and predict the class of the unlabeled one by their majority class. A vector of features describes the examples and the similarity between them is expressed by a distance function, usually the Euclidean distance. Thus, according to the vector of features and the distance function, the k closest instances to the unlabeled case are the k nearest neighbors used to classify it . The k-NN can be easily extended to regression. In this case, the target variable is numerical. When the target variable is unknown, the k-NN approach tries to find the k nearest neighbors within the set of inputs whose target value is known. The predicted target value is either the mean or the median. The k-NN regression can be applied to a univariate time series problem. In this case, the target of an example is a historical value of the time series, and its associated feature vector is described by the lagged values of the target . Figure 1 shows an example of one-step-ahead forecasting using 2-NN approach. The feature vector is composed by the two previous values of the target variable. We construct a feature vector based on the last two values in the time series to forecast the next period. Finally, the two closest feature vectors are determined and their target values are averaged to produce the forecast data point (red point).

\begin{figure}[htp]
\begin{center}
\includegraphics[width=6.7in,height=3in]{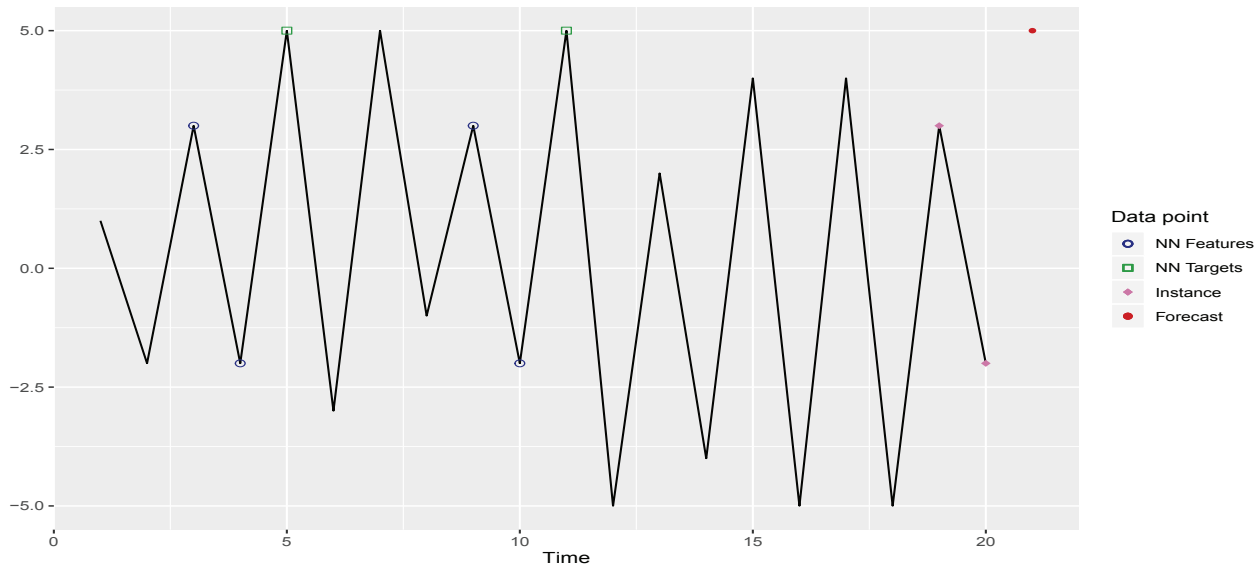}
\caption{\protect\parbox[t]{10cm}{one step ahead forecast with 2-NN regression.}}
\label{1:f1}
\end{center}
\end{figure}
\section{Preprocessing for time series forecasting}
 Feature preprocessing techniques in time series have a significant influence on the model's performance and forecasting accuracy. Therefore, they are essential in a forecasting model. For example, when dealing with a non-stationary time series, classical times series methods are applied to transform the original series to a stationary one. A time series is stationary if its statistical properties are all constant over time. It is one whose properties do not depend on the time at which the series is observed.  Several studies advocate the deseasonalization of time series for neural networks that meet the propriety of seasonality (Ahmed et al., 2010). That is, we increase the prediction accuracy by fitting the model on the transformed time series. In this section, we present and analyze some of well-known techniques used for time series preprocessing.

\subsection{Outliers }
 An outlier is a data point that differs significantly from other observations that can distort statistical results. The source of the outliers may be due to variability in the measurement or to experimental (data entry) error. If the cause can be attributed to a data entry error, the error is easily fixed.  It is important to suss out any potential outliers in our data. This can be performed visually via a scatterplot of the response versus each predictor by looking for any bivariate outliers or points that fall outside the general trend of the data. The type of the outliers in a time series are: additive outlier (AO), temporary change (TC), level shift (LS), and innovational outlier (IO) (Abraham et al., 2009). In the k-NN regression setting, an outlier has the potential to distort predictions and affect the accuracy( Martínez et al., 2019). There are many strategies for dealing with outliers in data. In this paper, we adopt the heuristic method described in Yan (2012), which is adopted by Martínez et al.(2019). That is, an observation is considered an outlier if its absolute value is four-time greater than the maximum of the absolute value of the median of the three consecutive points that are located before and after our observation . Particularly, a point is an outlier if the following condition is met:
\begin{equation}
\label{eq:1}
 |o_i| \geq 4 × \max\{|m_b |, |m_a|\},
\end{equation}
\noindent where $m_b= median(o_{i-3}, o_{i-2}, o_{i-1})$ and $m_a = median(o_{i+1}, o_{i+2}, o_{i+3})$. Once an observation is determined to be an outlier, we replace it by the average of the two adjacent observations that are located before and after the outlier. For  time series data, to detect and correct an outlier, we apply a method described  by Yan (2012): \\
\noindent\rule{\linewidth}{2pt}\smallskip

\noindent \textbf{Algorithm 1: } \textit{DAO(x)}\smallskip

\noindent\rule{\linewidth}{1pt}

1:\qquad \textbf{Input} : $x$ time series before outliers' adjustment

2:\qquad  \textbf{Output} : $x$ after outliers' adjustment

3:\qquad  $T\leftarrow length(x)$

4:\qquad  \textbf{for} $i=4$  \textbf{to} $T-3$ \textbf{step} 1  \textbf{do}

5:\qquad  \qquad \textbf{if} $|x_{i} |\geq 4*max( \mid median(x_{i-3}, x_{i-2}, x_{i-1})\mid,\mid median(x_{i+1}, x_{i+2}, x_{i+3})\mid)$

6:\qquad  \qquad \qquad \textbf{print} $i$

7:\qquad  \qquad \qquad $x_{i}\leftarrow \frac{1}{2} (x_{i-1}+x_{i+1} )$

8:\qquad  \qquad \textbf{end}\textbf{if}

9:\qquad  \textbf{end}\textbf{for}

10:\quad  \ \ \textbf{end}.\\
\noindent\rule{\linewidth}{2pt} \smallskip  \\

\subsection{ Stabilizing variance in time series }
 The application of the k-NN regression in a multiplicative time series is problematic. In  this case, it cannot predict values outside the range of the data because it predicts the average values in historical data (Martínez et al., 2019). A problem that is known as extrapolation.  In a multiplicative seasonal time series, extrapolation may result in biased estimates because the variance of the estimates varies over time. To stabilize the variance in this type of time series, we apply the box cox transformation (Box and Cox, 1964), which is defined as follows:
\begin{equation}
\label{eq:2}
boxcox(x_{t},\lambda)=  \left\{ \begin{array}{l}
\log(x_{t}) \ \ if \ \ \lambda=0 \\
\frac{x^{\lambda}_{t}-1}{\lambda} \ \ otherwise
\end{array} \right.
\end{equation}
\noindent where $\lambda\in \mathbb{R}$ is determined by maximizing the log-likelihood of the fitted linear model (Box and  Cox, 1964) or minimizing the coefficient of variation of subseries of x (Guerrero, 1993). We select $\lambda$ by applying the BoxCox.lambda function in R package "forecast" (Hyndman et al., 2019) where the Guerrero method is used as a default.

\subsection{Detrending a time series}
Implementing the k-NN regression to a time series data that exhibits a global trend may result in inaccurate and unprecise estimates (Martínez et al., 2019). Recall that k-NN predicts observation that lies inside the data range. To detrend a time series one may apply differencing method or Seasonal and Trend decomposition using Loess (STL) approach (Hyndman and  Athanasopoulos, 2018; Cleveland et al., 1990). Using the differencing method implies computing the differences between consecutive observations. That is, we work with transformed variable $w_{t} = \Delta^{d}(x_{t})=(1-L)^{d}(x_{t})$ where $L^{d} (x_{t})=x_{t-d}$\quad  for $ d\in \mathbb{N}$. In a k-NN regression setting, if the lags used as input variables are the set l, then the differences at lag min(l) should be taken (Martínez et al., 2019). In time series and given a target value, lags 1 to p are used as input variables when its p previous values are used as input variables  (Martínez et al., 2018). On the other hand, using the STL decomposition implies the decomposition of series as a combination of trend, seasonal, and remainder components.  Once we remove the trend by implementing the Holt's method, we forecast the seasonal and remainder components by the k-NN regression method. The forecast of the trend component, the seasonal component, and the remainder component are added to generate the final forecast. In this paper we implement the differencing approach to detrend the time series data.

\section{ Methodologies}
In this section, we present our methodologies, which apply the k-NN approach to time series forecasting. The first method uses classical parameters tuning in WNN (CPTO-WNN) and the second method uses fast parameters tuning in WNN (FPTO-WNN). The main idea of both methods is to select adequate parameters for a number of neighbors k and the window's length p (WNN) by using an optimization tool. Precisely, the time series is divided into a set of training and test sets. The WNN parameters are selected by minimizing the average accuracy measure, which is calculated over all the test data.
\subsection{CPTO-WNN method }

Given a time series $x = (x_{1},x_{2},...,x_{T})$, the problem consists of predicting the next values  $s_{T}=(x_{T+1},x_{T+2},...,x_{T+n})$.  Let $ss_{i} = (x_{i-p+1},x_{i-p+2},....,x_{i})$ be a feature vector and $s_{i}=(x_{i+1},x_{i+2},....,x_{i+n})$ be its associated target. For any  i=p, p+1,....,T and j=p,p+1,...,T, we can associate a distance as follows:
\begin{equation}
\label{eq:3}
dist(i,j)=dist(ss_{i},ss_{j}),
\end{equation}
\noindent where dist is the Euclidean distance. The idea of WNN used in CPTO-WNN method consists of the determination of a set of k nearest neighbors to  $ss_{T}$, which we define the neighbor set (NS) as:\\
 $NS= \{ set \ \ of \ \ ss_{T_{1}},ss_{T_{2}},....,ss_{T_{k}}  \ \ closest \ \ to \ \  ss_{T} \ \ | \ \ T_{j} \in \{p,p+1,....,T-n\}  \}$,
 in which $ss_{T_{j}}$  refers to the j-th neighbor to $ss_{T}$, in order of distance.\\
Note that NS is defined when $k \leq T-n-p+1  $, where p and k must satisfy the following condition :
\begin{equation}
\label{eq:4}
p+k \leq T-n+1.
\end{equation}
\noindent After we determine the neighbor set, we forecast $s_{T}$  using the weighted mean of the targets associated to the k nearest neighbors to  $ss_{T}$ :
\begin{equation}
\label{eq:9}
\widehat{s_{T}}  =\sum_{j=1}^{k}  \frac{ w_{j}}{\sum _{i=1}^{k}\mathds{1}(dist(T,T_{i})=dist(T,T_{j}))} \sum _{i=1}^{k}\mathds{1}(dist(T,T_{i})=dist(T,T_{j}))*s_{T_{i}},
\end{equation}
\noindent where $w_{j} \in [0,1]$ is a weighting factor related to the degree of closeness of $ss_{T_{j}}$  to $ss_{T}$  .
$\mathds{1}(x=y)$ represents the indicator function of event $x=y$ that takes the value 1 if the
assertion $x = y$  is true and 0 otherwise. In our work, we employ weights based on Rank Order Centroid (Barron, 1992), which are defined as follows:
\begin{equation}
\label{eq:10}
w_{j}= \frac{1}{k} \sum _{i=j}^{k} \frac{1}{i} , 
\end{equation}
\noindent for \ \ j=1,2,....,k. That is, $s_{T_{j}}$  is close to $s_{T}$ when $ss_{T_{j}}$  is close to $ss_{T}$ .

\subsubsection{Selection of the model parameters}

 To find the optimal value k in the nearest neighbors and the optimal window's length p, we implement a method inspired by the cross-validation method (Hyndman, 2016). The main difference of this method compared to the traditional cross-validation method is in the length of the test dataset. In the original method, there is a series of test dataset that consists of  a single observation, while the modified approach uses a series of test dataset that consists of a set of observations. The advantage of using the modified cross-validation method is that it is computationally less expensive compared to the traditional one. Precisely, we use training and test sets to select the optimal values of k and p by minimizing the Mean Absolute Percentage Error (MAPE). The MAPE is a measure prediction accuracy of a forecasting method and is defined as follows:\\
$MAPE = \frac{100}{n}*\sum\limits_{t=1}^{n}|\frac{y_{t}-\hat{y}_{t} }{y_{t}}|$ Where $y_{t}$ is the actual observed value, $\hat{y}_{t}$ is the predicted value for the time point t, and n is the number of observations. The modified time series cross validation method is defined as follows. Let $x=\{x_{1},x_{2},...,x_{T} \}$ be an univariate time series and $E=\{E_{1},....,E_{I}\}$  represents a set of $I$ training sets of $x$. For $i\in\{1,...,I\}$, $E_{i}=\{x_{1},x_{2},...,x_{T-i*n}\}$  is the i-th training set used to forecast the i-th test set  $s_{T-n \ast i}=(x_{T-i*n+1},x_{T-i*n+2},....,x_{T-i*n+n})$.  For $p\in \mathbb{N}^{*}$  and $k\in \mathbb{N}^{*}$, we calculate the accuracy of the test set  $s_{T-n\ast i}$ :\\
$MAPE_{i}(p,k)=  \frac{100}{n}*\sum\limits_{j=1}^{n}|\frac{\hat{x}_{T-i*n+j}-x_{T-i*n+j}}{x_{T-i*n+j}}|$ where $\hat{x}_{T-i*n+j}$ is calculated  by  (5). To measure the global accuracy of the model, we consider the average accuracy calculated over all test sets, which is  denoted by $MAPE^{*}(p,k)=\frac{1}{I} \sum\limits_{i=1}^{I} MAPE_{i}(p,k)$.
The values of $p$ and $k$ must satisfy (4) for each $E_{i}$. That is,
\begin{equation}
\label{eq:11}
p+k \leq T-n \ast I-n +1.
\end{equation}
\noindent The optimal values of $p$ and $k$ denoted by $p^{*},k^{*}$, respectively, are the values that minimize the $MAPE^{*}(p,k)$:
$$(p^{*},k^{*} )=argmin MAPE^{*}(p,k).$$

\subsubsection{CPTO-WNN's algorithms}
 In this section, we layout two algorithms that we implement to forecast time series. The first algorithm uses one given training set, a number of neighbors k, a number of length's window p, and a number of predictions to be made. The second algorithm implements the first algorithm and searches the optimal values of k and p given I training sets to forecast n future values.\\
\noindent\rule{\linewidth}{2pt}\smallskip

\noindent \textbf{Algorithm 2:} \textit{WNNtraining1(x,k,p,n)}\smallskip

\noindent\rule{\linewidth}{1pt}

1:\qquad \textbf{Input} : the training set $(x)$, the number  of neighbors $(k)$,

\qquad \qquad \qquad   \ \ the window's length $(p)$,  the number of predictions to be made $(n)$

2:\qquad \textbf{Output} : Mean Absolute Percentage Error  (MAPE)

3:\qquad $D$, $Dm$, s, f, b $\leftarrow$ empty vectors

4:\qquad T$\leftarrow$ length(x)

5:\qquad   $g_{t} \leftarrow boxcox\{x_{t}$  where $t= \{1,2,....,T-n\} , \lambda \}$

6:\qquad  $h_{t}\leftarrow g_{t+1}-g_{t}$

7:\qquad  \textbf{for} $j = 1$ \textbf{to} $T-2*n-p$ \quad  \textbf{step}  1  \textbf{do}

8:\qquad  \qquad    $D_{j} \leftarrow dist( \{h_{T-n-p},h_{T-n-p+1},....,h_{T-n-1} \} , \{h_{j},h_{j+1},....,h_{j+p-1} \} )$

9:\qquad  \textbf{endfor}

10:\quad \ \ $Dm$ $\leftarrow$ sort $D_{j}$ in ascending order

11:\quad \ \ $NS\leftarrow\{j$ where $D_{j}=Dm_{o}$ with $o =1,....,k\}$

12:\quad \ \ $Y\leftarrow g_{T-n}$

13:\quad \ \ \textbf{for} $c=1$ \textbf{to}  $n$ \textbf{step}  1 \textbf{do} 

14:\quad \ \ \qquad     \qquad \qquad  $s_{c} \leftarrow \sum_{j=1}^{k}    \frac{w_{j}}{\sum\limits_{i\in NS}\mathds{1}(D_{i}=Dm_{j})} \sum\limits_{i\in NS}\mathds{1}(D_{i}=Dm_{j})*h_{i+p+(c-1)}$

15:\quad \ \ \qquad   \qquad \qquad  $b_{c}\leftarrow s_{c}+Y$

16:\quad \ \ \qquad    \qquad \qquad    $Y \leftarrow b_{c}$

17:\quad \ \ \qquad   \qquad \qquad    $f_{c}\leftarrow boxcox^{-1}(b_{c},\lambda)$

18:\quad \ \ \textbf{endfor}

19:\quad \ \ return $MAPE( \{f_{1},f_{2},....,f_{n} \},\{x_{T-n+1},x_{T-n+2},....,x_{T} \})$

20:\quad \ \ \textbf{end}.

\noindent\rule{\linewidth}{2pt} \smallskip  \\ \\

\noindent\rule{\linewidth}{2pt}\smallskip

\noindent \textbf{Algorithm 3:} \textit{WNNoptimization1(x,w,k,I,n)}\smallskip

\noindent\rule{\linewidth}{1pt}

1:\quad\textbf{Input}: the time series to forecast$(x)$, the maximal window's length entered by the user$(w)$, 

\qquad \qquad \quad the maximal number of neighbors entered by the user $(k)$,  

\qquad \qquad \quad the number of training sets  entered by the user $(I)$,

\qquad \qquad \quad the number of future values to be predicted $(n)$

2:\quad \textbf{Output} :$p^{*}$ and $k^{*}$ minimizing $ MAPE^{*}$
 
\qquad \qquad \quad (the average mean of MAPE calculated over all test sets)

3:\quad \ \ $m$ $\leftarrow$ empty raw

4:\quad \ \ $l \leftarrow 10000$

5:\quad \ \ $T\leftarrow length(x)$

6:\quad \ \ \textbf{for} $c = 1$ \textbf{to}  $k$ \textbf{step}  1 \textbf{do} 

7:\quad \ \ \quad  \textbf{for} $p = 1$ \textbf{to}  $w$ \textbf{step}  1 \textbf{do} 

8:\quad \ \ \qquad    \textbf{for} $i = 1$ \textbf{to}  $I$ \textbf{step}  1 \textbf{do} 

9:\quad \ \ \qquad \quad          $m_{i}\leftarrow  WNNtraining1(\{x_{1},x_{2},....,x_{T-(i-1)*n} \},c,p,n)$

10:\quad \ \ \qquad   \textbf{endfor}

11:\quad \ \ \qquad   \textbf{if} \ \ $\frac{1}{I}\sum_{i=1}^{I}m_{i} < l$ \textbf{do} 

12:\quad \ \ \qquad \quad     $l \leftarrow \frac{1}{I}\sum_{i=1}^{I}m_{i}$

13:\quad \ \ \qquad \quad    print $p$

14:\quad \ \ \qquad \quad   print $c$

15:\quad \ \ \qquad\quad  print $l$

16:\quad \ \ \qquad \textbf{endif}

17:\quad \ \ \quad \textbf{endfor}

18:\quad \ \ \textbf{endfor}

19:\quad \ \ \textbf{end}.

\noindent\rule{\linewidth}{2pt} \smallskip  \\
\subsection{FPTO-WNN}
 FPTO-WNN method is an approach that we derive to overcome the high computation cost in the CPTO-WNN method. The WNN approach used in the FPTO-WNN works as follows. Given a time series $x=(x_{1},x_{2},....,x_{T} )$, the goal consists of predicting the next values $s_{T}=(x_{T+1},x_{T+2},....,x_{T+n})$.
Let $s_{i}=(x_{i+1},x_{i+2},....,x_{i+n})\in \mathbb{R}^{n}$ be the target of the feature vector $ss_{i}=(s_{i-n*p},s_{i-n*(p-1)},...,s_{i-n})$. For  any $i=T,T-n,.....,T-n\ast c$ and $j=T,T-n,.....,T-n\ast c$ , a distance is defined as follows:
\begin{equation}
\label{eq:13}
 dist(i,j)=dist(ss_{i},ss_{j}),
\end{equation}
\noindent where dist is the Euclidean distance and c is the last integer such that $T-n*c \geq n*p$. That is, when $T-n*c$ is strictly less than $n*p, ss_{T-n*c}$ is not defined.\\
\noindent First, the method consists of identifying the k nearest neighbors of
$$ss_{T}= (s_{T-n*p},s_{T-n*(p-1)},...,s_{T-n})= (x_{T-p*n+1},x_{T-p*n+2},...,x_{T} ),$$ where k and p must be determined. The neighborhood in this context is measured according to (8), which leads to the following neighbor set:\\ 
$ NS=\{ set \ \ of \ \ ss_{T_{1}},ss_{T_{2}},....,ss_{T_{k}}  \ \ closest \ \ to \ \ ss_{T} \mid T_{j} \in \{ T-n,T-2\ast n,...,T-n\ast c\} \}$, in which $ss_{T_{1}}$  and $ss_{T_{k}}$ refer to the first and the k-th neighbors, respectively.\\
Note that NS is defined when $k\leq c \Rightarrow  n*k \leq n \ast c \Rightarrow T-n\ast k\geq T-n\ast c $. Since $ T-n\ast k\geq T-n\ast c$ and $T-n\ast c\geq n\ast p$, $T-n\ast k$  should be greater than $n*p$, and then p and k must satisfy the following condition:
\begin{equation}
\label{eq:5}
p+k \leq \frac{T }{n}.
\end{equation}
\noindent Once we determine the neighbor set, we forecast $s_{T}$ using the weighted mean:
$$\widehat{s_{T}}  =\sum_{j=1}^{k}  \frac{ w_{j}}{\sum _{i=1}^{k}\mathds{1}(dist(T,T_{i})=dist(T,T_{j}))} \sum _{i=1}^{k}\mathds{1}(dist(T,T_{i})=dist(T,T_{j}))*s_{T_{i}},$$
\noindent where $w_{j}$ is calculated by formula (6).
\subsubsection{Selection of  model parameters}
To find the optimal values for $p$ and $k$, we implement the modified cross validation method as explained above where $p$ and $k$ meet the condition in (9) for each training set
 $E_{1\leq i\leq I} = (x_{1},x_{2},...,x_{T-i*n})$. That is,
\begin{equation}
\label{eq:6}
p+k \leq \frac{T-n*I}{n}.
\end{equation}
\subsubsection{FPTO-WNN's algorithms }
 In this section, we layout two algorithms that we implement to forecast time series using the FPTO-WNN approach. The main difference compared to the CPTO-WNN algorithm lies in the neighbors' search set.\\

\noindent\rule{\linewidth}{2pt}\smallskip

\noindent \textbf{Algorithm 4:} \textit{WNNtraining2(x,k,p,n)}\smallskip

\noindent\rule{\linewidth}{1pt} 

1:\quad \ \ \textbf{Input}:The training set$(x)$, the number  of neighbors $(k)$,  the window's length  $(p)$,

\qquad \qquad \quad the number of predictions to be made $(n)$

2:\quad \ \ \textbf{Output }: Mean Absolute Percentage Error  ($MAPE$)

3:\quad  \ \ $D$,$Dm$, s, f, b $\leftarrow$ empty vectors

4:\quad \ \ $T\leftarrow length(x)$

5:\quad  \ \ $g_{t} \leftarrow  boxcox\{x_{t} \quad   where \quad  t= \{1,2,....,T-n\},\lambda \} $

6:\quad  \ \ $h_{t} \leftarrow g_{t+1}-g_{t}$

7:\quad  \ \ \textbf{for} $j = T-(p+2)*n$  \textbf{to}  1 \textbf{step} -n  \textbf{do} 

8:\qquad \ \ \qquad  $D_{j}\leftarrow dist( \{h_{T-n*(p+1)},h_{T-n*(p+1)+1},...,h_{T-n-1} \}  ,   \{h_{j},h_{j+1},...,h_{j+p*n-1} \})$

9:\quad \textbf{endfor}

10:\qquad  $Dm$ $\leftarrow $Sort $D_{j}$ in ascending order

11:\qquad  $NS\leftarrow\{j \quad where \quad D_{j}=Dm_{o} \quad with \quad o =1,....,k\}$

12:\qquad  $Y\leftarrow g_{T-n}$

13:\quad \ \ \textbf{for} $c=1$ \textbf{to} n \textbf{step} 1 \textbf{do} 

14:\quad \ \ \qquad  \qquad   $s_{c}\leftarrow \sum_{j=1}^{k}    \frac{w_{j}}{\sum\limits_{i\in NS}\mathds{1}(D_{i}=Dm_{j})} \sum\limits_{i\in NS}\mathds{1}(D_{i}=Dm_{j}) *h_{i+n*p+(c-1)}$

15:\quad \ \ \qquad   \qquad  $b_{c}\leftarrow s_{c}+Y$

16:\quad \ \ \qquad  \qquad  $Y\leftarrow b_{c}$

17:\quad \ \ \qquad  \qquad   $f_{c}\leftarrow (boxcox)^{-1}(b_{c},\lambda)$

18:\quad \ \ \textbf{endfor}

19:\quad \ \ return $MAPE( \{f_{1},f_{2},....,f_{n} \},\{x_{T-n+1},x_{T-n+2},....,x_{T} \})$

20:\quad \ \ \textbf{end.}

\noindent\rule{\linewidth}{2pt} \smallskip  \\ \\


\noindent\rule{\linewidth}{2pt}\smallskip

\noindent \textbf{Algorithm 5:} \textit{WNNoptimization2(x,n,k,w,I)}\smallskip

\noindent\rule{\linewidth}{1pt} 

1:\quad \textbf{Input} : the time series to forecast $(x)$, the number of future values to be predicted $(n)$,

\qquad \qquad \quad   the maximal number of neighbors entered by the user $(k)$,

\qquad \qquad \quad   the maximal window's length entered by the user $(w)$,

\qquad \qquad \quad  the number of training sets entered by the user $(I)$

2:\quad \textbf{Output} : $p^{*}$ and $k^{*}$ minimizing $MAPE^{*}$

\qquad \qquad \qquad \small (the average mean of MAPE calculated over all test sets)

3:\quad  \ \ $m$ $\leftarrow $ empty raw

4:\quad \ \ $l \leftarrow 10000$

5:\quad \ \ $T \leftarrow length(x)$

6:\quad \ \ \textbf{for} $c = 1$ \textbf{to} $k$ \textbf{step} 1 \textbf{do}

7:\quad \ \ \quad  \textbf{for} $p = 1$ \textbf{to} $w$ \textbf{step} 1 \textbf{do}

8:\quad \ \ \qquad     \textbf{for} $i = 1$ \textbf{to} $I$ \textbf{step} 1 \textbf{do}

9:\quad \ \ \qquad \quad  $m_{i} \leftarrow WNNtraining2(\{x_{1},x_{2},....,x_{T-(i-1)*n} \},c,p,n)$

10:\quad \ \ \qquad    \textbf{endfor}

11:\quad \ \   \qquad  \textbf{if} $\frac{1}{I}\sum_{i=1}^{I}m_{i} < l$ \textbf{do}

12:\quad \ \   \qquad \quad   $l \leftarrow \frac{1}{I}\sum_{i=1}^{I}m_{i}$

13:\quad  \ \    \qquad \quad  \textbf{print} $p$

14:\quad \ \      \qquad \quad \textbf{print} $c$

15:\quad  \ \     \qquad \quad  \textbf{print} $l$

16:\quad  \ \    \qquad \textbf{endif}

17:\quad \ \ \qquad \textbf{endfor}

18:\quad \ \ \quad \textbf{endfor}

19:\quad \ \ \textbf{end.}

\noindent\rule{\linewidth}{2pt} \smallskip \\

Note that when $n = 1$, WNNoptimization1 and WNNoptimization2 are the same.


\section{ Real data analysis }
This section presents an evaluation of the two new methods derived in section 4 using two real-time series data. The data are the Retail and Food Services Sales in the USA and Milk Production in the UK. We implement our methods in R. To detect and adjust outliers, a  preprocessing of data is performed by applying the DAO algorithm presented above. Then, we implement the WNNoptimization1 and WNNoptimization2 algorithms to the preprocessed time series where we set k = 10, w =15, and the size of the test set be equal to $30\%$ of the available data: $n*I=0.3*T$. $T$ is the length of the time series. The objective is that for each n (=1,2,....,10) and based on conditions (7) and (10), we select the best values for $k^{*}\in\{1,2,....,10\}$ and $p^{*}\in\{1,2,....,15\}$ in order to fit an optimal WNN model.

\subsection{ Example1: Retail and Food Services Sales in the USA }

 We use the retail and food services sales data in the United States Census Bureau (2020).  The time series include 338 points, range from January 1992 to February 2020, and are delineated by month. Figure 2 depicts the monthly sales for Retail and Food Services Sales in the US.\\
\begin{figure}[htp]
  \begin{center}
 \includegraphics[width=6.7in,height=3in]{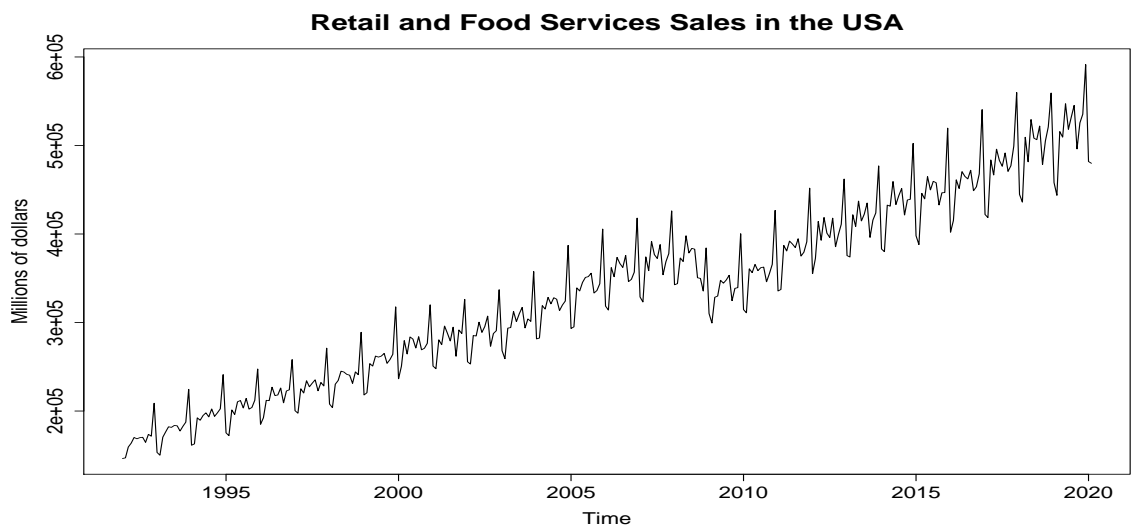}
  \caption{Monthly sales for series Retail and Food Services Sales in the USA.}\label{hist1}
\end{center}
\end{figure}
The implementation of the DAO algorithm shows the absence of the outliers. Table 1 presents the optimal values for p and k using both WNNoptimization1 and WNNoptimization2 algorithms, along with their MAPE and the computation time in seconds (CT). We use a horizon $n \in \{1,2,....,10\}$. It is evident that the WNNoptimization1 method outperforms the WNNoptimization2 method. However, the first algorithm is computationally expensive with an average computation cost of 14 min. Whereas, the second algorithm has an average computation cost of 8 min.  We select n =10 as it provides less computation time (361.07s). In addition, the second method provides a good accuracy and a lower computation time across all horizons. Figure 3 presents the prediction based on CPTO-WNN method for n=10. That is, we compare the predicted values by our model to the actual values for the period ranging from January 2011 to February 2020 . Our method provides results almost identical to the observed one, with similar patterns and high accuracy. Similarly, Figure 4 depicts the prediction based on the FPTO-WNN method when n=7 for the period ranging from June 2011 to February 2020. Similarly, it is evident that our predictions are so close to the actual values. Also, our methods provide high accuracy and similar patterns as the observed one.
\begin{table}[htp]
\begin{center}
\begin{tabular}{c c|c c c c|c c c c c c}
  \hline
  \multicolumn{2}{c}{Method} & \multicolumn{4}{|c|}{WNNoptimization1} & \multicolumn{4}{c}{WNNoptimization2} \\ \hline
Horizon n & I  & $p^*$ & $k^*$ & MAPE     & CT(s)   & $p^*$ & $k^*$ & MAPE   & CT(s)    \\ \hline
1	      & 102 & 14  &  8  & 1.308579  & 2612.29 & 14  & 8   & 1.308579  & 2612.29 \\
2         & 51 & 6   &  9  & 1.290372 & 1343.20  & 3  & 9   & 1.290372 & 787.08 \\
3         & 34 & 8  &  8 & 1.614844 & 931.95  &  2  & 8  & 1.62293 & 428.52 \\
4	      & 26 &  6  &   8 & 1.459892 & 727.31  &  2  & 8  & 1.488859 & 303.11 \\
5         & 21 & 7  &  10 & 1.637205  & 604.04  &  6  &  2  & 1.970321 & 239.04 \\
6         & 17 &  9  &  10 & 1.688256 & 506.04  &  1  & 10  & 1.69961 & 197.56 \\
7         & 15 &  8  &   8 & 1.285574 & 460.39  & 2  &  3  &  2.371874 & 181.45 \\
8         & 13 &  4  &   3 & 1.65241 & 411.81  &  7  &  10  &  1.736769 & 163.74 \\
9         & 12 &  8  &  10 & 1.371099 & 388.66  &  1  &  6  &  1.588464 & 158.56\\
10        & 11 &  6  &   10 & 1.50652 & 361.07  & 3  & 4  &  2.015993 & 122.2\\
  \hline
\end{tabular}
\caption{Selection of the optimal parameters ($p^*$  and $k^*$) by each method.}
\end{center}
\end{table}

\begin{figure}[htp]
  \begin{center}
 \includegraphics[width=6.7in,height=3in]{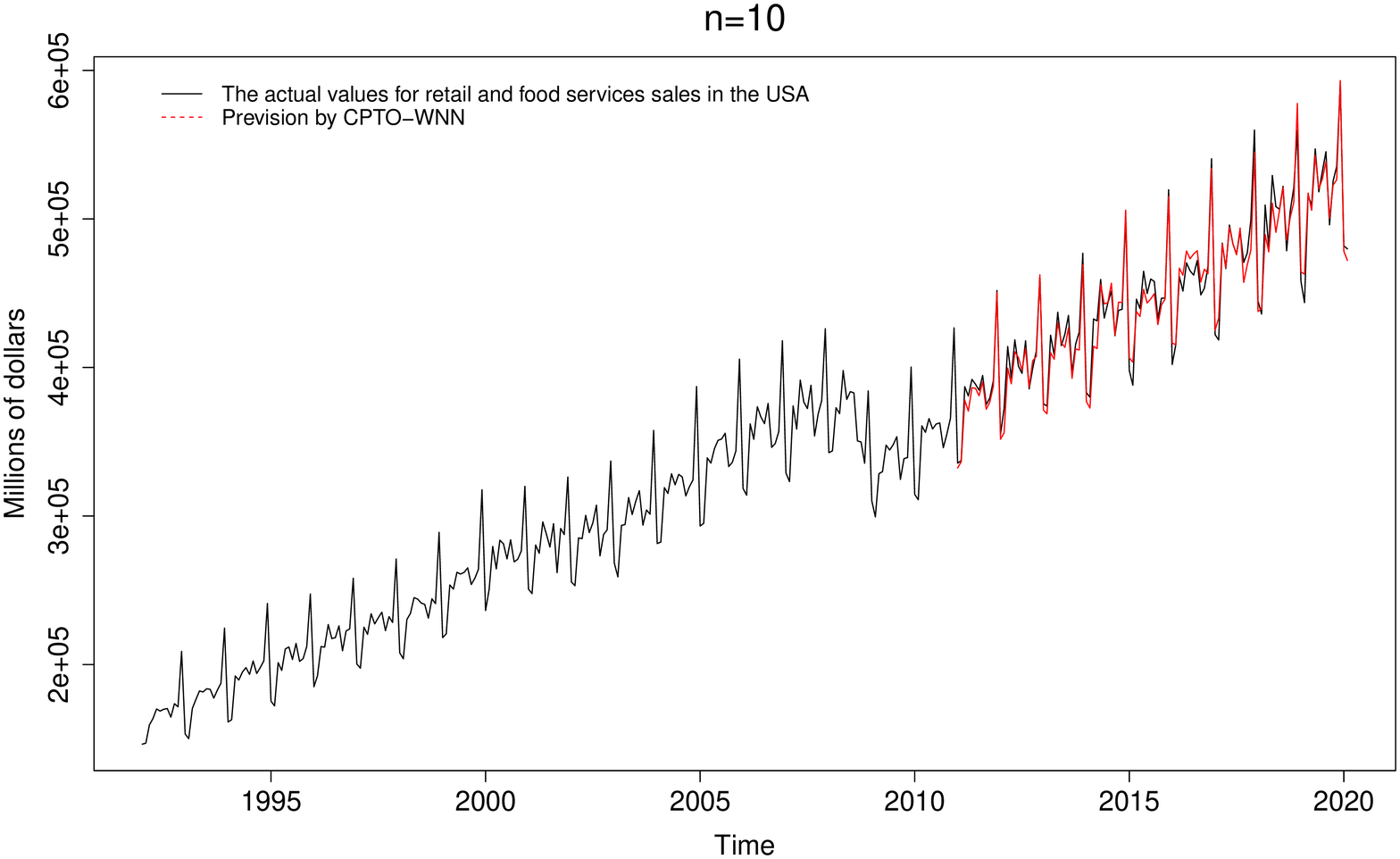}
  \caption{The observations and the predictions for Retail and Food Services Sales in USA using the CPTO-WNN method for n = 10}\label{hist2}
\end{center}
\end{figure}

\begin{figure}[htp]
  \begin{center}
 \includegraphics[width=6.7in,height=3in]{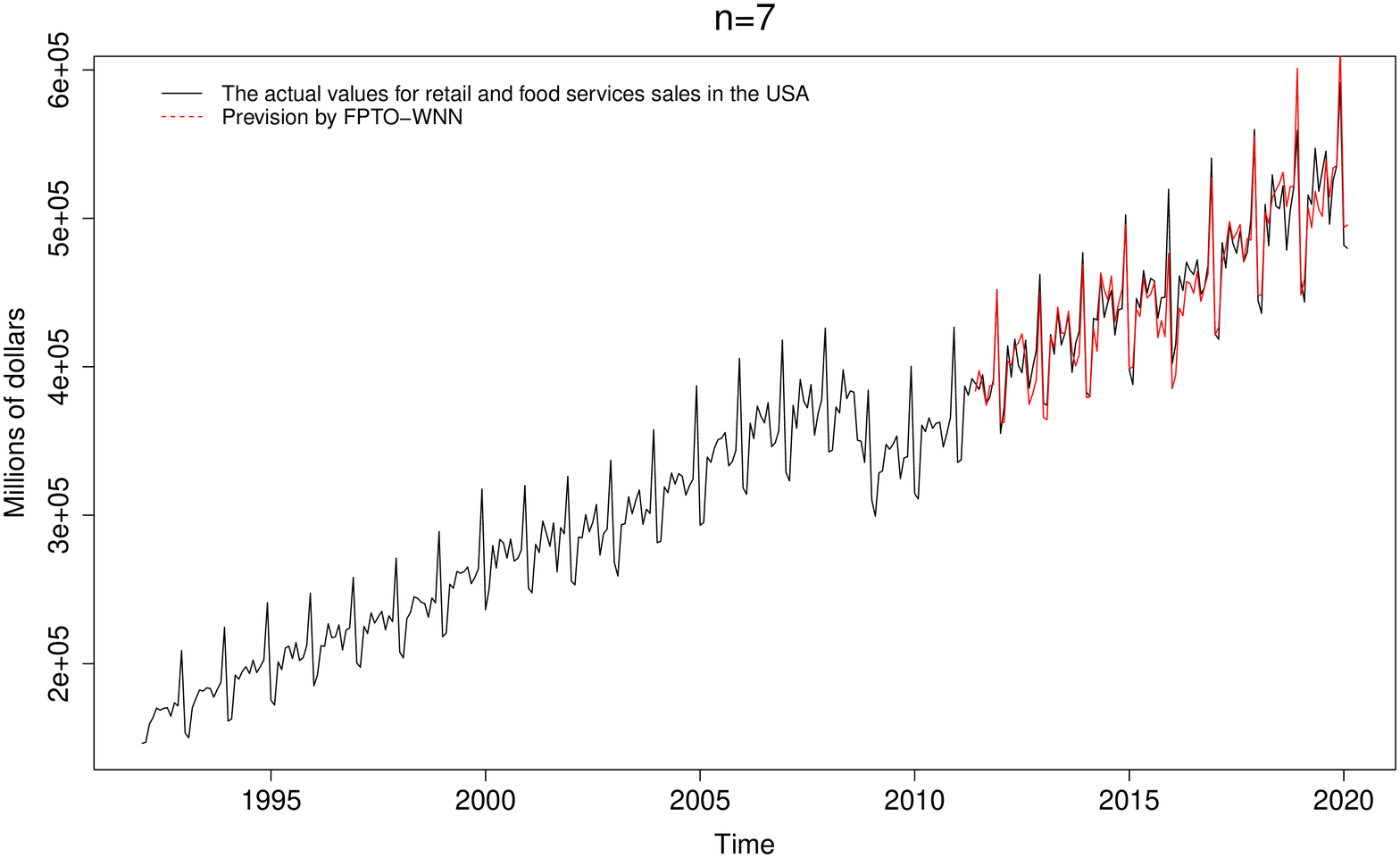}
  \caption{The observations and the predictions for Retail and Food Services Sales in USA using the FPTO-WNN method for n=7}\label{hist3}
\end{center}
\end{figure}

\subsection{Milk Production in the UK }

In this example, we use the cow's milk production in the UK  (Eurostat, 2020). The time series include 411 points, range from January 1986 to March 2020, and are delineated by month. Figure 5 depicts the monthly cow's milk production in thousand tonnes in UK.
\begin{figure}[htp]
  \begin{center}
 \includegraphics[width=7in,height=4in]{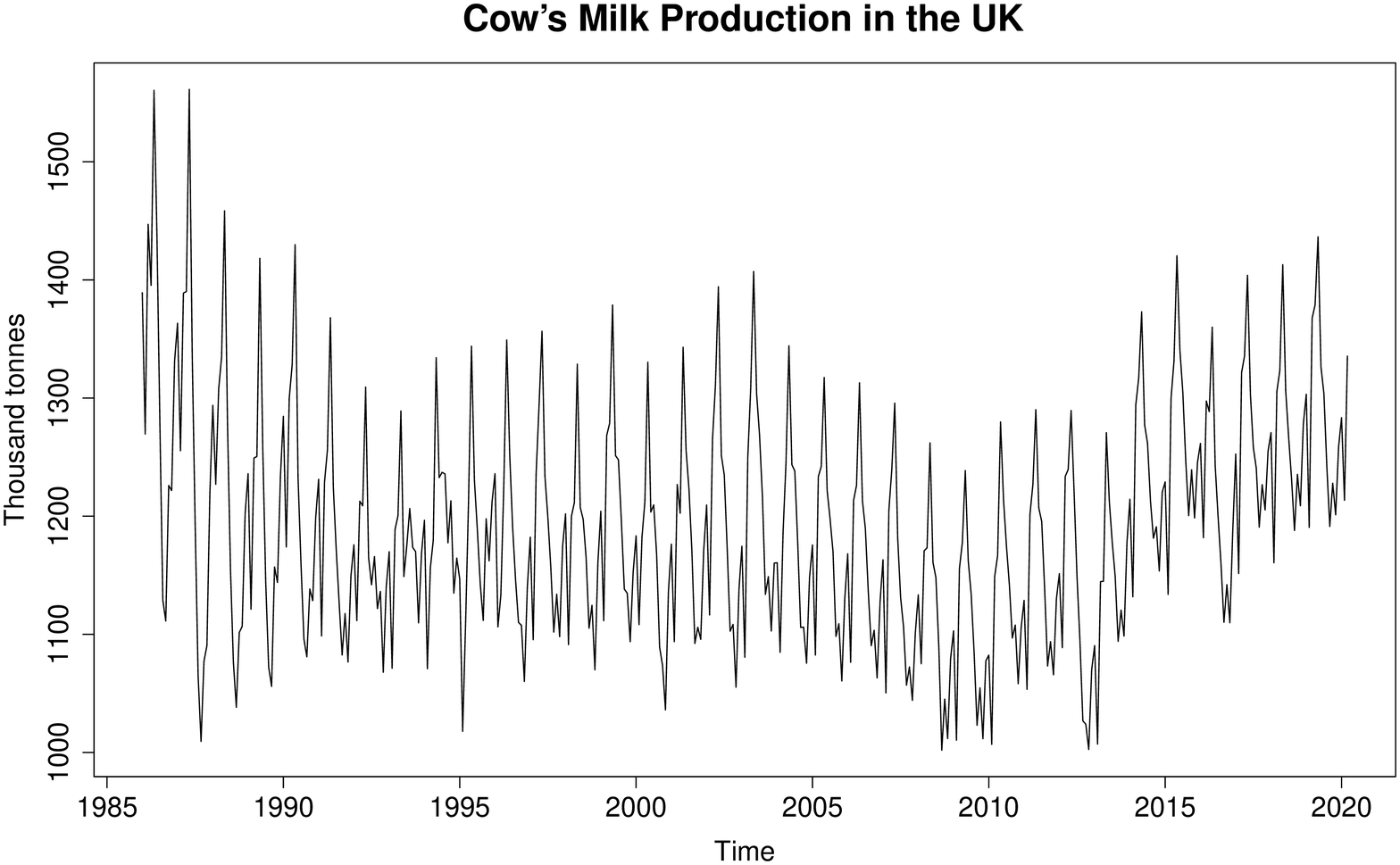}
  \caption{Cow's Milk Production in the UK}\label{hist4}
\end{center}
\end{figure}
Exploratory data analysis and the implementation of the DAO algorithm show the absence of outliers. Table 2 presents the optimal values for p and k using both WNNoptimization1 and WNNoptimization2 algorithms, along with their MAPE and the computation time in seconds (CT). For the horizons n = 3,4,5,7,9 the WNNoptimization1 algorithm provides slight high accuracy compared to the WNNoptimization2 algorithm. However, WNNoptimization1 algorithm is computationally expensive with an average computation cost of 24min. Whereas, the second algorithm has an average computation cost of 16 min. To sum up, one can conclude that the FPTO-WNN method outperforms the CPTO-WNN method. Figure 6 illustrates the predictions based on CPTO-WNN method for $n\in\{1,3,4,6\}$. We compare the predicted values to the actual values for the last $I*n+n$ observations. It is evident that our models provide results almost identical to the observed one, with similar patterns and high accuracy. Similarly, Figure 7 depicts the predictions based on FPTO-WNN method for  $n\in\{2,3,4,10\}$. The method provides results with high accuracy and with similar patterns compared to the observed data.
\begin{table}[htp]
\begin{center}
\begin{tabular}{c c|c c c c|c c c c c c }
  \hline
  \multicolumn{2}{c}{Method} & \multicolumn{4}{|c|}{WNNoptimization1} & \multicolumn{4}{c}{WNNoptimization2} \\  \hline
Horizon n & I  & $p^*$ & $k^*$ & MAPE     & CT(s)   & $p^*$ & $k^*$ & MAPE   & CT(s)    \\ \hline
1	      & 124 & 15  &  10  & 1.276587  & 4164.06 & 15  & 10   & 1.276587  & 4164.06 \\
2         & 62 & 10  &  9  & 1.566065 & 2223.36  & 5  & 9   & 1.566065 & 1464.13 \\
3         & 42 & 13  &  10 & 1.671981 & 1622.85  &  8  & 10  & 1.682875 & 894.99 \\
4	      & 31 &  9  &   6 & 2.124234 & 1248.34  &  2  & 8  & 2.179696 & 652.42 \\
5         & 25 & 9  &  10 & 1.998537  & 1064.10  &  4  &  7  & 2.062267 & 549.47 \\
6         & 21 &  13 &  10 & 2.343873 & 936.25 &  5  & 10  & 2.171279 & 483.06 \\
7         & 18 &  12 &   2 & 2.662673 & 844.61  & 1  &  7  &  2.903861 & 442.40 \\
8         & 16 &  8  &   6 & 2.576981 & 779.78  &  14  &  5  &  2.464982 & 422.14 \\
9         & 14 &  4  &  5 & 2.083085 & 718.93 &  2  &  10  &  2.203154 & 401.25\\
10        & 13 &  15  &   10 & 3.069087 & 700.86  & 3  & 10  &  2.729576 & 395.26\\ \hline
\end{tabular}
\caption{ Selection of the optimal parameters ($p^{*}$  and $k^{*}$) by each method.}
\end{center}
\end{table}

\begin{figure}[htp]
  \begin{center}
 \includegraphics[width=6.7in,height=3.5in]{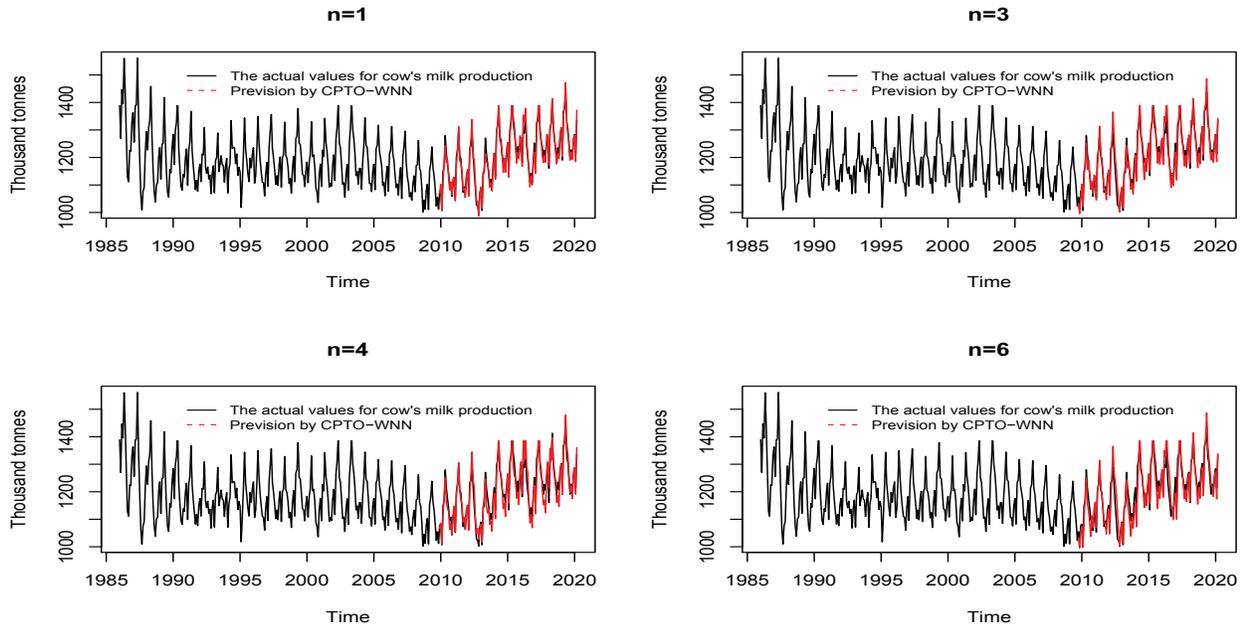}
  \caption{The observations and the predictions for Cow's Milk Production in the UK using the using the CPTO-WNN method  for n=1, 3, 4  and 6.}\label{hist5}
\end{center}
\end{figure}

\begin{figure}[htp]
  \begin{center}
 \includegraphics[width=6.7in,height=3.5in]{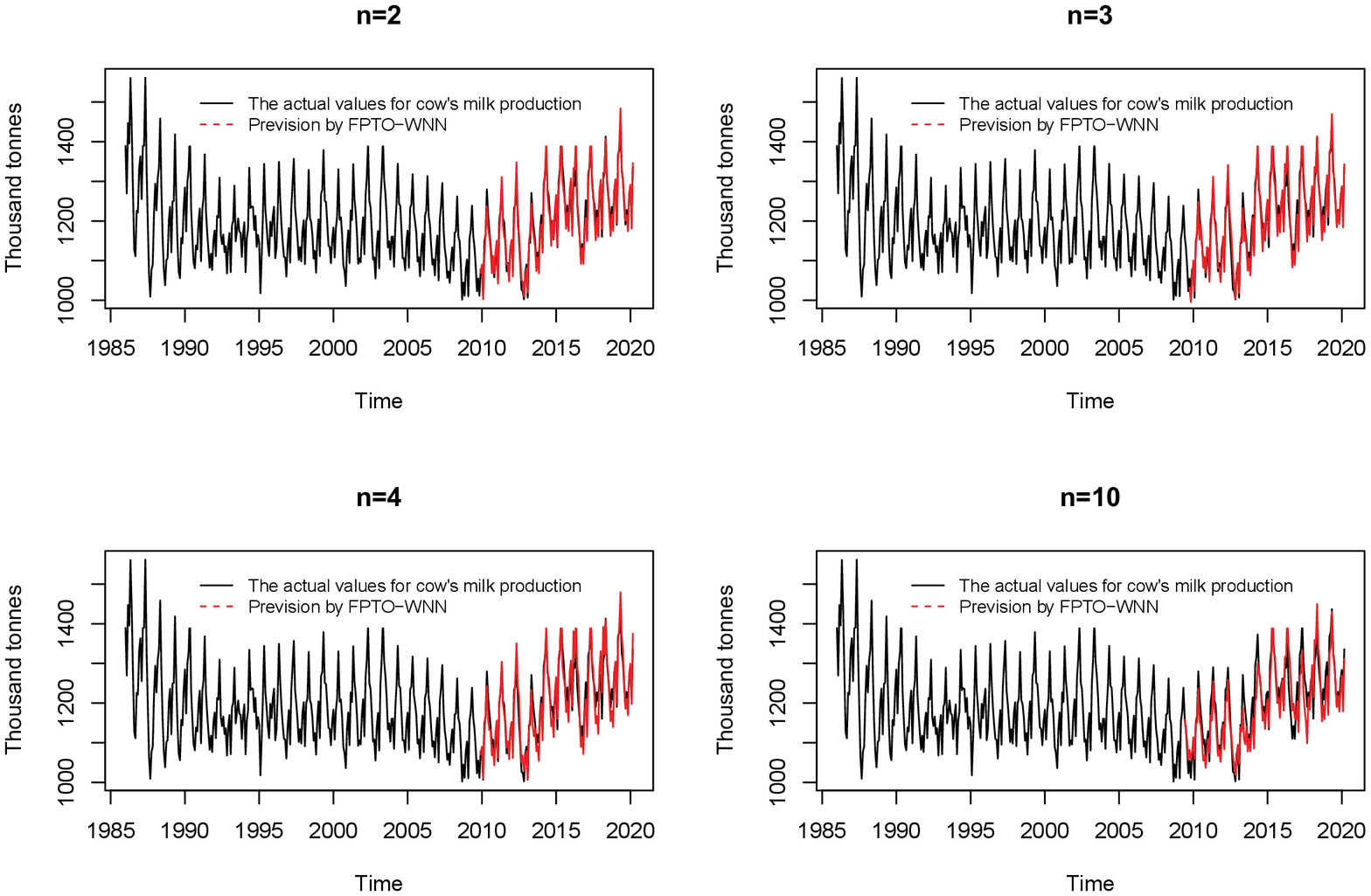}
  \caption{The observations and the predictions for Cow's Milk Production in the UK using the FPTO-WNN method for n=2, 3, 4 and 10.}\label{hist6}
\end{center}
\end{figure}


\section{Comparison with Other Techniques }

 This section compares our derived methods with the some classical approaches to time series forecasting. Precisely, we compared them to SARIMA, Holt-Winters and Exponential smoothing state space model (ETS) (Hyndman and Athanasopoulos, 2018). We partition the data into training sets and test sets as described in section 4. For each training set, we fit the models where we use the functions "auto.arima", "ets" and "hw" from the R package "forecast" (Hyndman R et al., 2019) to fit the classical models. Then, based on the models that we fit on a training dataset, we forecast the corresponding test dataset, which consists of the n observations  that had been realized just after the observations that form the training dataset. Finally, the forecast accuracy is computed by averaging an accuracy measure over the test sets. We use the MAPE function.\\
\noindent Tables 3 and 4 present the MAPE results for different methods using the Retail and Food Services Sales data  in the USA and the Cow's Milk Production data in the U.K, respectively. Similarly, Figures 8 and 9 depict methods comparison based on MAPE using the Retail and Food Services Sales data in the USA and the Cow's Milk Production data in the U.K, respectively. From Table 3 and Figure 8, one can see that there is no one model to rule them all, and the performance definitely depends on the selection of the horizon. Plus, all methods provide good accuracy. In addition, the CPTO-WNN method outperforms FPTO-WNN, SARIMA, Holt-Winters, and ETS methods when n=2,7,9 and 10, and outperforms the FPTO-WNN method across all horizons. Also, the Holt Winters method fail when n=1,2.\\

\begin{table}
\begin{center}
{\scriptsize
\begin{tabular}{c c|c|c|c|c|c}
  \hline
\multicolumn{2}{c|}{Methods} & CPTO-WNN  &  FPTO-WNN & SARIMA & Holt Winters & ETS  \\
Horizons & I  &  \qquad  &   \qquad&  \qquad &  \qquad &   \\ \hline
1        &  102  &  1.308579  &  1.308579  &  1.29761  &  Optimization failure  &  1.370033  \\
2        &  51  &  1.290372  &  1.290372  &  1.332103  &  Optimization failure &  1.31375  \\
3        &  34  &  1.614844  &  1.62293 &  1.415163  &  1.427527  &  1.420891  \\
4        &  26  &  1.459892  &  1.488859  &  1.366595  &  1.381188  &  1.350499  \\
5        &  21 & 1.637205 &  1.970321  &  1.49651  &  1.512568  &  1.515745  \\
6        &  17 & 1.688256  &  1.69961  &  1.713453  &  1.621776  &  1.552982  \\
7        &  15 & 1.285574  &  2.371874  &  1.542372  &  1.399539  &  1.396673  \\
8        &  13 & 1.65241  &  1.736769  &  1.574424  &  1.659944  &  1.622978  \\
9        &  12 & 1.371099  &  1.588464  &  1.643077  &  1.536242  &  1.519837  \\
10       &  11  & 1.50652  &  2.015993  &  1.67685  &  1.51898  &  1.571653  \\ \hline
\end{tabular}
}
\caption{Methods comparison based on MAPE using the Retail and Food Services Sales data  in the USA using different horizons.}
\end{center}
\end{table}
\begin{figure}[htp]
  \begin{center}
 \includegraphics[width=6.7in,height=3.5in]{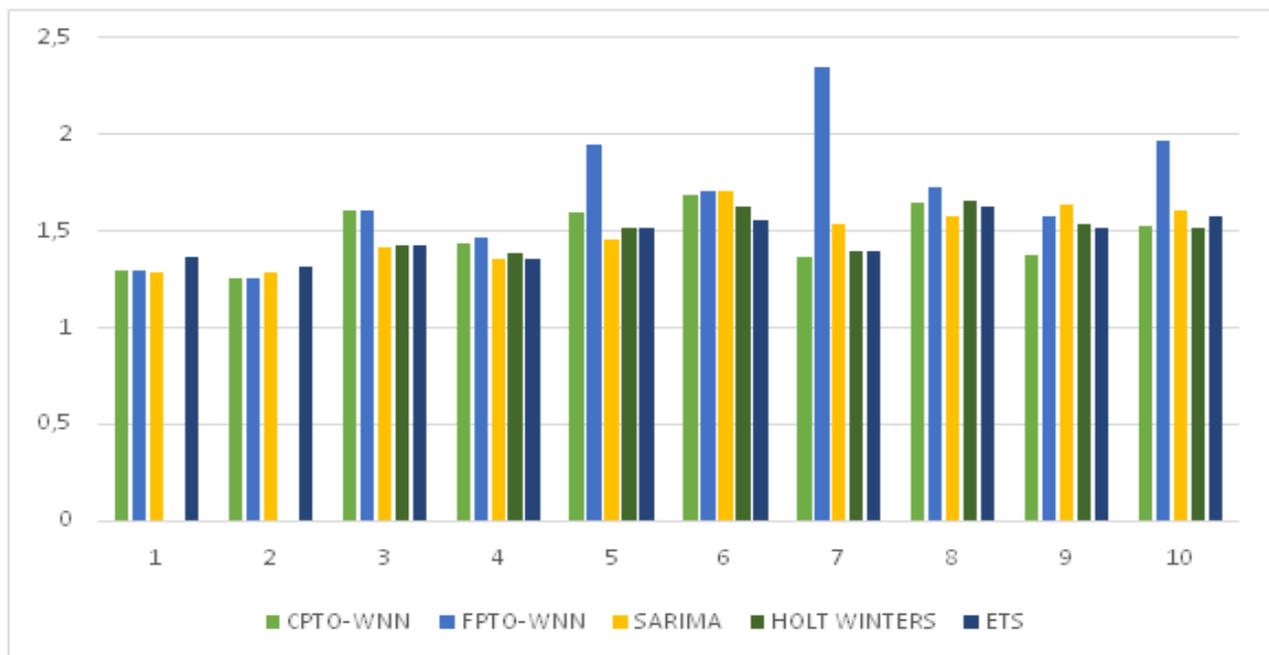}
  \caption{Methods comparison based on MAPE using the Retail and Food Services Sales data  in the USA using different horizons.}\label{hist7}
\end{center}
\end{figure}
From Table 4 and Figure 9, it is evident that for all methods as we increase the number of the horizon, the accuracy decreases. However, the CPTO-WNN and FPTO-WNN methods greatly outperform SARIMA, Holt-Winters, and ETS method across all horizons. In addition, the FPTO-WNN approach is computationally the less expensive. That is, in this case the FPTO-WNN method rules them all.\\
\begin{table}
\begin{center}
{\scriptsize
\begin{tabular}{c c|c|c|c|c|c}
  \hline
\multicolumn{2}{c|}{Methods} & CPTO-WNN  &  FPTO-WNN & SARIMA & Holt Winters & ETS  \\
Horizons & I  &  \qquad  &   \qquad&  \qquad &  \qquad &   \\ \hline
1  &  124  &  1.276587  &  1.276587  &  1.352952  &  1.651371  &  1.349542  \\
2  &  62  &  1.566065  &  1.566065  &  1.671056  &  1.846018  &  1.624575 \\
3  &  42  &  1.671981  &  1.682875  &  1.958765  &  2.053901  &  1.809076  \\
4   &  31  &  2.124234  &  2.179696  &  2.348748  & 2.360477  &  2.265423  \\
5  &  25 & 1.998537  &  2.062267  &  2.639499  &  2.300733  &  2.460435  \\
6  &  21 & 2.343873  &  2.171279  &  2.508519  &  2.675902  &  2.451997  \\
7  &  18 & 2.662673  &  2.903861  &  3.167989  &  2.884389  &  2.832336  \\
8  &  16 & 2.576981  &  2.464982  &  2.964856  &  2.951236  &  2.475475  \\
9  &  14 & 2.083085  &  2.203154  &  2.789552 &  2.510556  &  2.485949  \\
10 &  13  &  3.069087  &  2.729576  &  3.378276  &  3.213982  &  3.093789  \\ \hline
\end{tabular}
}
\caption{Methods comparison based on MAPE using the Cow's Milk Production data in the U.K using different horizons.}
\end{center}
\end{table}
\begin{figure}[htp]
  \begin{center}
 \includegraphics[width=6.7in,height=3.5in]{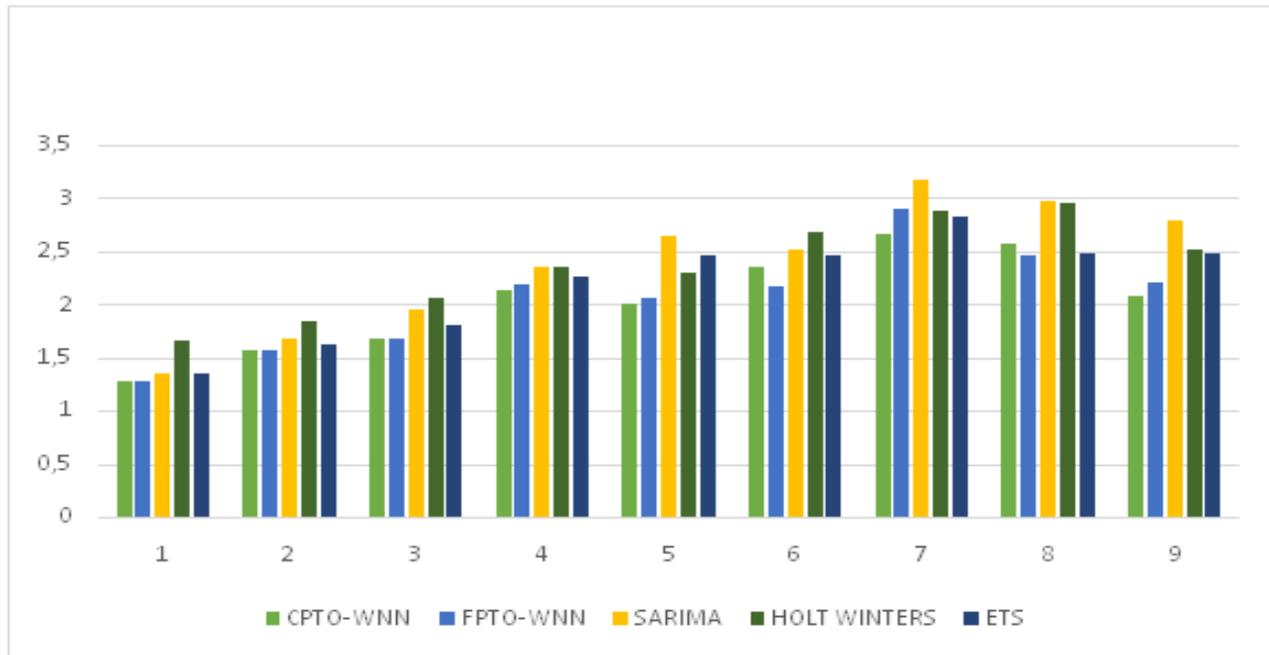}
  \caption{Methods comparison based on MAPE using the Cow's Milk Production data in the U.K using different horizons.}\label{hist8}
\end{center}
\end{figure}

\newpage
\section{Conclusion: }
In this paper, we derived two methods for time series forecasting. Both methods use a modified cross validation to Weighted Nearest Neighbors model. We built two automatic algorithms that we named WNNoptimization1 and WNNoptimization2. For each iteration, WNNoptimization1 compared the most recent subsequence with every possible subsequence from the past of the same length. However, this raises concerns about the computational time mostly when we have too many iterations. To overcome this, we introduced the WNNoptimization2 method that reduces the neighbors search set. Real data examples show that our approach outperforms the classical approaches in terms of efficiency and accuracy.\\

When the dimension of the data is large, our algorithms might raise concerns about the computational complexity. To overcome this, one can implement the algorithms in the framework of Apache Spark. Apache Spark is a unified analytics engine for big data processing. In addition, we want to introduce a Bayesian approach to the k-NN regression, which will allow us to introduce interpretability to the results and to incorporate external information and/or historical information through  priors. We will explore these suggestions in future work.\\ \\

\noindent \textbf{Data Availability Statement}\\
For our analysis, the Retail and Food Services Sales in the USA data that support the finding of this study are obtained from the United States Census Bureau Database, available at:\\
https://www.census.gov/retail/index.html.\\
Cow's Milk Production in the UK are provided by Eurostat, which are available at:\\ http://appsso.eurostat.ec.europa.eu/nui/show.do?dataset=apro\_mk\_colm\&lang=en.



\newpage

\makeatletter
\renewcommand{\@biblabel}[1]{}
\makeatother




\end{document}